# Multistable rippling of graphene on SiC: A Density Functional Theory study

*Tommaso Cavallucci, and Valentina Tozzini\**

NEST, Scuola Normale Superiore and Istituto Nanoscienze-Cnr, Piazza San Silvestro 12, 56127, Pisa, Italy

**ABSTRACT** Graphene monolayer grown by Si evaporation from the 0001 surface of SiC displays a moiré pattern of corrugation whose structure is ambiguous: different measurements and theoretical studies show either protruding bumps surrounded by valleys, or, reversely, wells surrounded by crests. Here we address the fine structure of monolayer graphene on SiC by means of Density Functional Theory, using a model including the full symmetry of the system and the substrate (1648 atoms) and therefore realistically reproducing the experimental sample. We find that accurate treatment of the vdW interactions between monolayer and the underlying substrate-bound buffer layer is crucial in stabilizing one or the opposite corrugation pattern, which explain the different results and measurement available in the literature. Our study indicates that at low temperature a state more closely following the topography of the underneath buffer layer is stabilized, while others are metastable. Since environmental conditions (e.g. temperature or doping) can influence the vdW forces and reduce the energy differences, this system is prone to externally driven switching between different (opposite) corrugation states. In turn, corrugation is related to local reactivity and to electronic properties of graphene. This opens to potentially interesting applications in nano-electronics or tailored graphene chemical functionalization.

## 1. INTRODUCTION

Thermal decomposition of SiC is a technique that allows growing high purity wafer-scale graphene samples, directly on an insulating substrate[1], useful for advanced applications[2]. The method is based on silicon evaporation from SiC surfaces, and consequent reconstruction of excess carbon in graphene-like layers[3]. In particular, the evaporation from 6H-SiC(0001) produces, in the first stage, the so-called "buffer layer"[4], namely a honeycomb carbon lattice partially covalently bound to the underlying Si atoms. STM measurements reveal that the buffer layer has a moiré pattern of corrugation, whose periodicity is 13×13 with respect to the graphene lattice[3] (or 6√3×6√3R30 with respect to SiC lattice, super-cell contoured in red in Figure 1). The corrugation pattern is often roughly described as a set of crests approximately following a honeycomb super-lattice (schematized by white lines in Figure 1A) whose periodicity is "quasi" 6×6[1,3-6] with respect to SiC (or 4√3×4√3R30 with respect to graphene, yellow in Figure 1A). As Si is further evaporated, another carbon layer forms underneath the previous one, replacing the buffer layer, while the top one detaches and acquires the electronic properties of graphene[5]. This is called the "monolayer" and is also corrugated. Its corrugation however, is less pronounced and has, therefore, a less clear pattern, showing either a pattern of type A, similar to that of the buffer, but more smeared (Figure 1A, pattern A), or "inverted", displaying hills located at the ~6×6 triangular lattice sites, separated by valleys (Fig 1A, pattern B). In some of the STM measurements[3,6], especially those on samples at low temperature[7], pattern A seems prevalent, while in others a mixture of the two[5] with a prevalence of B[8] can be guessed. We observe that, from the purely geometrical point of view, pattern B can be obtained from A by a reflection with respect to the graphene plane plus a translation of 1/3 along the *x* lattice vector of the 6√3×6√3R30 supercell. We remark that, while in pattern A the depressions and protrusion of buffer- and mono-layer coincide, in pattern B the corrugation of the monolayer is partially mismatched, having intrusions in part superimposed to protrusions of the buffer.

The corrugation pattern of graphene on SiC was also addressed in a number of theoretical studies based on Density Functional Theory (DFT). It was shown that the moiré pattern with approximate symmetry could be obtained in the isolated suspended sheet by isotropic lateral compression of the 4√3×4√3R30 super-cell[9]. In this case, the symmetry is superimposed, and states A and B are completely degenerate, being related by simple symmetry operations. Conversely, bare lateral compression of the super-cell with the symmetry of the real experimental sample (13×13) does not return any experimental-like pattern in the suspended sheet, indicating that the interactions with buffer layer have a fundamental role in the stabilization the specific curvature pattern. Several DFT studies are available of graphene on SiC,[10,11,12,13]. However, most of them consider too small super-cells to address the corrugation issue. To our knowledge, only a few represent the system with the full 13×13 cell including several layers of the substrate, the buffer layer and the monolayer [14,15,16] differing by the calculation setup: ref [14] and [16] use the Local Density Approximation (LDA)[17] for the exchange and correlation energy functional, but with different basis sets for the orbitals expansion, while ref [15] uses the General Gradient Approximation (GGA)[18]. In [14] and [15], the buffer layer corrugation pattern is in agreement with the experimental one (in [16] corrugation is not reported). However, they diverge in the representation of the monolayer, being of type A in ref [14] and of type B in ref [15]. In summary, nor the experimental neither the theoretical studies solve the ambiguity between A-like and B-like patterns.

The aim of this work is to specifically address the fine structure and relative stability of the different corrugation patterns of monolayers. Besides the interest in clarifying the actual structure of graphene on SiC, the local curvature of the sheets has recently emerged as a technologically interesting property. For instance, since the reactivity of C atoms is strongly correlated to the local out of plane deformation[19], the curvature control was proposed as a possible mechanism for loading and release hydrogen in graphene based storage devices [20,2]. Other applications might involve tailoring the hydrogen decoration[21] of graphene for hybrid graphane/graphene systems for nano-electronics[22]. In addition, because curvature is related to reactivity in general, not only to hydrogen chemisorption, rippling manipulation might be used to control amount and location of any chemical species or molecule on graphene[23] and be used to tailor its functionalization. Therefore, controlling the transition between two structural patterns might have a number of interesting applications.

In the next section, we describe in detail the calculation setup. Different electron density functionals are used, with and without van der Waals corrections. A presentation of the results follows, with their discussion. A summary and possible developments of this work are reported in the Conclusions section.

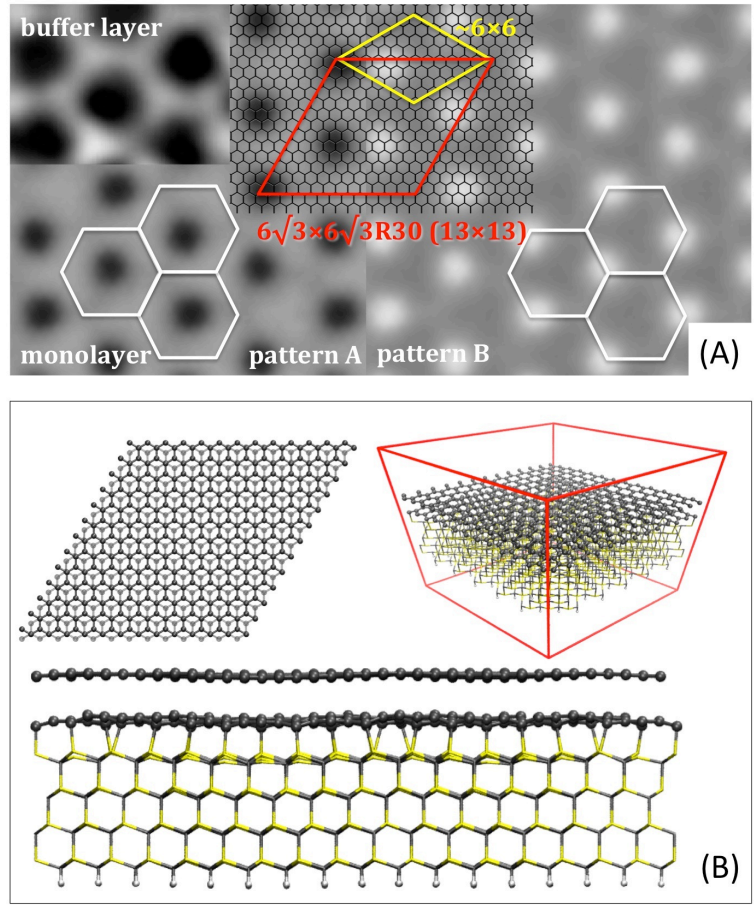

**Figure 1.** (a) Schematic representation of the curvature patterns of buffer and monolayer: pattern A = wells separated by crests, pattern B = hills separated by valleys (lighter areas are protruding; images are representative artwork mimicking super-atomic resolution STM images). The graphene lattice is superimposed (in black) and the two reference supercells (6√3×6√3R30 or 13×13, and ~6×6) are reported in red and yellow respectively. White lines roughly represent the buffer layer crests pattern within the simplified 6×6 symmetry. (b): top, perspective and side view of the super cell of the model system for calculations. In the perspective view the super-cell simulation box boundaries are reported in red.

## 2. METHODS

The model system we used for the main calculations is represented in Figure 1(b). It consists of the 6√3×6√3R30 supercell one saturated with H atoms. The system is periodic in the three directions, included z, and the c parameter is large enough to leave ~16 Å of empty space between periodic images. We performed structural optimization both for the system with buffer layer only (1310 atoms) and for the complete system (with monolayer over buffer layer, 1648 atoms). Starting configuration are obtained subsequently placing flat 13×13 graphene sheets over the SiC surface and relaxing. The buffer layer is placed at ~2Å from the SiC, and the monolayer at ~3.4Å from the buffer in AB stacking, after relaxation of the latter. The relaxations adjust distances and refines stacking. Different relaxation runs were preformed starting from flat or differently rippled monolayers and at different distances, as specified further on (and in the SI,Table S.IV). Smaller model systems representing monolayer, bilayer and graphite with different interlayer distance were considered for test calculations to evaluate the inter-layer interaction energy. In these cases the minimal unit cell is considered, including two atoms for monolayer and four atoms for bilayer and graphite, and with variable c lattice parameter. Brillouin zone sampling was commensurate to the size of the unit cell: we used up to 24×24×2 k points distributed with the Monkhorst and Pack scheme[24] for smaller systems and the Γ point for the large ones. Gaussian smearing of energy depending on the cell was used for the integration over the Brillouin zone. The BFGS quasi-Newton algorithm was used for local minima search[25].

Calculations were performed using Rappe-Rabe-Kaxiras-Joannopoulos ultrasoft (RRKJUS) pseudopotentials[26] and a plane wave basis set with cutoff set at 30 Ry (density cutoff to 300 Ry). Our working hypothesis is that the bufferlayer-monolayer interactions have a fundamental role in the effect under study. Therefore, we used the combination of Perdew-Burke-Ernzerhof (PBE) exchange and correlation functional[27] with the empirical dispersion correction of Grimme (hereafter "PBE-D2"[28]), known to accurately reproduce the inter-layer distances and van der Waal (vdW) energies of graphite[29,30]. However, for comparison, we additionally considered calculation setups similar to that of refs [14] and [15], namely LDA[17] and PBE without vdW corrections (hereafter "PBE"). Cell parameters and all other details for the whole list of performed calculations are reported in Tables S.I and S.II and S.III of the supporting information file. All calculations were performed with Quantum Espresso (QE5.0.1[31]). Test calculations were addressed on local small clusters and multi-core workstations, while the full model systems were treated on the IBM Blue Gene/Q (FERMI@CINECA).

## 3. RESULTS

In order to assess the performances of the different calculations setup especially in determining the interlayer distances and energies, we first performed a set of calculations for graphite and bilayers with different inter-layer distances (reported in detail in the SI, section S.2). Our tests indicate that LDA performs well on the interlayer distance but underestimates of more than 50% the vdW interlayer energies. The underestimation of vdW energy is even larger, about one order of magnitude, with PBE setup, which also overestimate the interlayer distance of more than 10% (details on these calculations are reported in the SI, Fig S.1 and Table S.II). While comparing well with previous results[29,30], these indicate that the interlayer interactions are very sensitive to the calculation setup. Therefore, not only the accurate choice of functionals and corrections is essential to realistically represent a system, but also, the comparison between different calculations might bring significant information for real situations in which the interlayer interaction is changed by environmental conditions. Consequently, we performed calculations with the three mentioned setups also for the supported system.

### 3.1 Buffer layer structure

The optimized structure of SiC+buffer layer obtained with PBE-D2 is shown in Fig 2. The pattern of corrugation displays crests defining an irregular hexagonal tessellation of the plane. The vertices of the hexagon are "hot spots" of protrusion, represented in colors in the figure. The examination of the z-profiles along three main symmetry directions (the inset plot) indicates a maximum z displacement of less than 1Å, as measured from the lower C atoms located near the center of the "hexagonal tiles" of the tessellation to the hot spots. These low C atoms are bound to the Si of the SiC substrate, as shown by the analysis of the slices of the electronic density (snapshot profiled in colors), displaying quite elongated Si-C bonds in correspondence of the centers of the tiles. The profile analysis also shows that the "hot spots" are located at different height, but the difference of their protrusion is very small (of the order of 0.01Å). The protrusion profiles shown in Fig 2 are in agreement with the theoretical ones reported in ref [15] and with STM measurements[5].

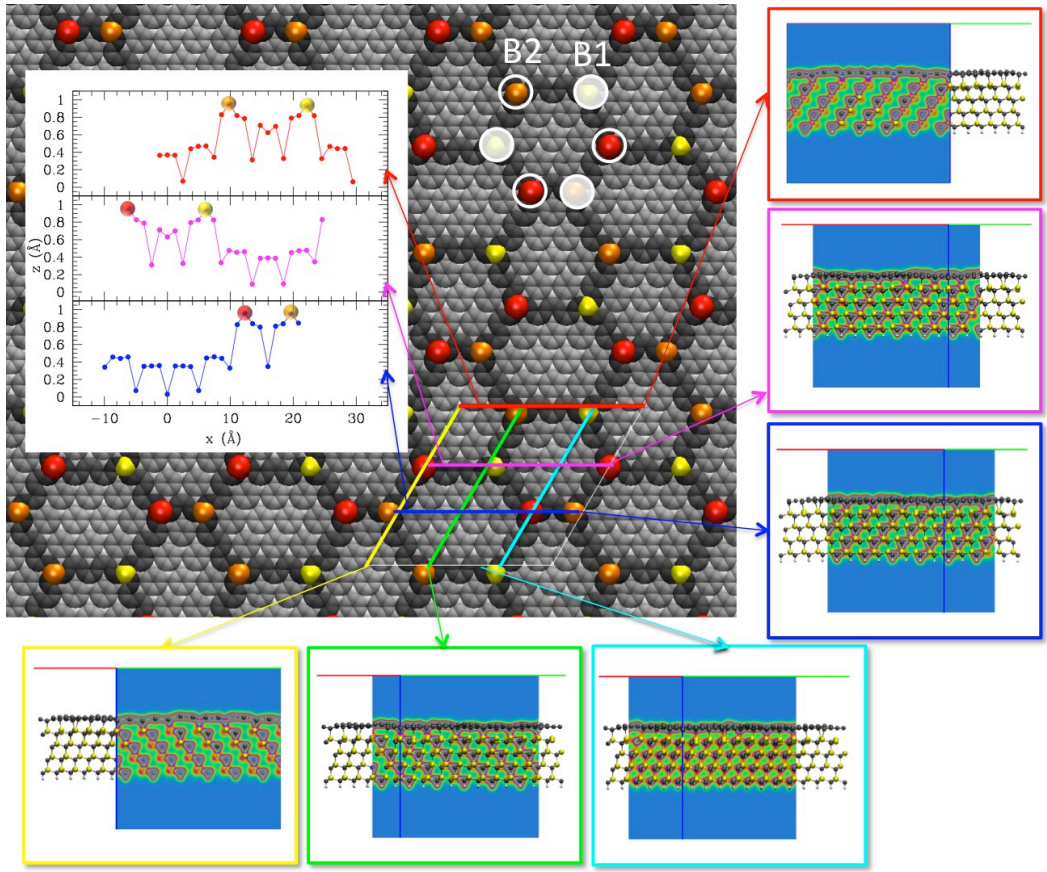

**Figure 2**. Structure of bufferlayer+SiC optimized with the PBE-D2 setup. In the background the buffer layer van der Waals surface is shown, colored according to the z coordinate (protruding atoms in darker grey). The atoms located uppermost ("hot spots") are highlighted in color, protrusion increasing from yellow to red (the very uppermost). They are located at the vertices of two non equivalent irregular triangular sublattices, named B1 and B2 (filled and empty circles respectively). The colored lines reported in the unit supercell represent vertical sections at which a volume slice of the electronic density is evaluated and reported in the plots on the right and at the bottom, contoured with colors corresponding to the lines. The color coding for density is: light blue for null density, green-yellow-orange-red for increasing density. The ball&stick structures are superimposed to the volume slice. The z protrusion profiles of the buffer layer only are also reported as a function of the x direction for the red, magenta and blue lines in the inset plot. Hot spots are also reported colored as in the structure.

The structure of the buffer layer seems quite insensitive to external conditions: the change in the buffer z profile induced by adding of the monolayer is of the order of 0.1Å, and the one due to change of functional around ~0.3Å (LDA flatter, PBE more protruding, PBE-D2 intermediate – details reported in section S.2.3 of the SI).

## 3.2 Monolayers structure

Conversely, the monolayer structure is strongly dependent on the simulation setup. The full system optimization with PBE-D2 reveals that the monolayer follows the same pattern of corrugation of the buffer layer, although with more smeared crests (Fig 3, central column), while the optimization with PBE returns a monolayer whose corrugation displays hills separated by valleys (Fig 3, right column), though with less pronounced curvature. In other words, PBE-D2 calculation produces state A, while PBE produces state B. Because PBE-D2 and PBE setups differ only by the vdW interactions (underestimated in PBE), these results clearly indicate that the relative stabilization of the two patterns is driven by the strength of the buffer-monolayer interaction. In addition, these results resolve discrepancies between different calculations in the literature: PBE setup stabilizes the system in a B-type curvature pattern, while stronger buffer-monolayer interactions (as our PBE-D2 or even LDA setup as in ref [14]) stabilize the system in A-type curvature pattern, which more closely follows the buffer.

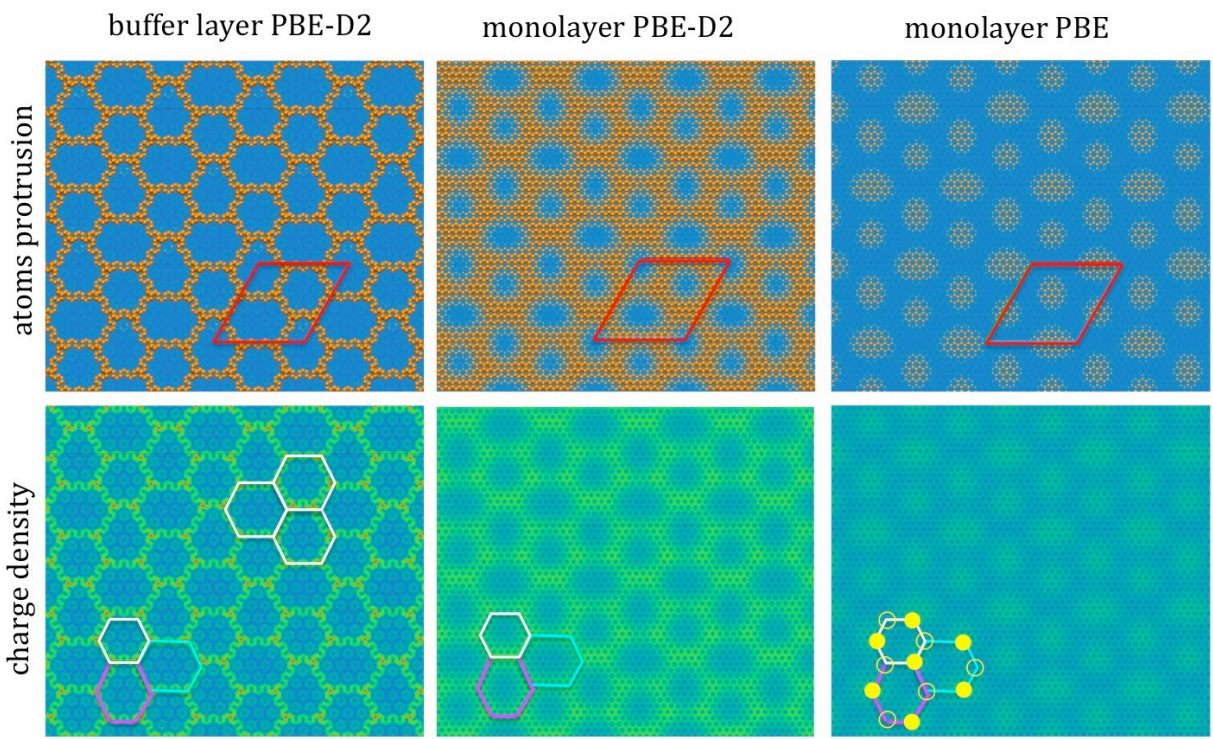

**Figure 3.** Representation of the uppermost layer corrugation in the optimized structures of the system with buffer layer only (with PBE-D2 setup), and with buffer+monolayer (PBE-D2 and PBE setup). In the first row, the protruding C atoms are represented in orange, by cutting the representation of the vdW surface of the layer with an horizontal blue plane (the part of the representation under the plane is not visible). In the second row, the electronic charge density is represented on an horizontal volume slice (light colors = higher density), giving an idea of the corresponding STM image. 13×13 supercell boundaries are reported in red, the schematic representation of crests is reported on the charge density representation (second row). In the one of monolayer-PBE, the non-equivalent sub-lattice sites of the crests pattern are indicated as filled or empty circles.

In order to better clarify this effect, we performed additional optimization runs starting from A- and B-like states. As already observed, the crests follow the edges of a tessellation of the plane with three kinds of irregular hexagons (in colors in the charge density representation of the buffer layer, Fig 3, bottom left image), fitting the " 13×13 supercell (reported in red). The regular hexagon pattern sometimes implied in the description of graphene on SiC, following the "quasi" 6×6 symmetry is only approximate (in white). Due to breaking of the hexagonal symmetry, the two sublattices of the tessellation (represented in yellow empty and filled dots in the bottom right of Fig 3) are not equivalent. Therefore, while there is only one possible A-type corrugation pattern matching with the buffer layer, two different B-type corrugation patterns exist, with the hills located on each of the two different sub-lattices of the crests pattern (i.e. empty or filled yellow dots). We named the one obtained from PBE optimization (reported in Fig 3, right column) as B1, and we also performed additional relaxations putting the system initially in the other one (B2) and in A, using also PBE and LDA setups.

Optimized structures are reported in Fig 4. The top view of the monolayer colored according to the z coordinate (red=protruding, blue=intruding) clearly put in evidence the differences between states A and Bs (columns): A state shows blue spots surrounded by red contours and B states show red spots surrounded by blue. Only slight differences between different calculations setups (rows) are visible. A and B curvature-like patterns of the monolayers can be roughly described as follows: A=wells located within the centers of the buffer hexagons and crests located over the buffer crests; B1= hills located over the B1 sublattice of hot spots of the buffer, separated by valleys; B2 = hills located over hot spots of the buffer layer separated by valleys; as said, specifically, B1 and B2 have hills located over one of the two non equivalent sublattices of hot spots (see fig 2). The buffer structural pattern (same as in Fig 2 and 3) is reported in black and white, to show the alignment with the corrugation of the monolayers. In all cases, the A state (blue and magenta) roughly follow the protrusion pattern of the buffer, though with less pronounced corrugation. The B1 and B2 states, conversely, have corrugation pattern partially displaced with respect to the buffer.

The z profiles of the buffer and monolayer are also reported in Fig 4 (bottom right plot, color coding of the lines is reported within the plot), and shows marked differences between different simulation setups: the buffer-monolayer distance for PBE is ~4.4Å, ~1Å larger than in the PBE-D2 and LDA case (~3.3Å and ~3.4Å). For comparison, the experimental value evaluated at room temperature can be estimated at ~3.6Å[32]. In addition, the average monolayer curvature level is nearly one half in PBE case with respect to the other two.

### 3.3 Relative stability of the states

Different corrugation states appear along different optimization runs started from different configurations (see the SI table S.IV for the full list of runs) as final or intermediate states. Selected optimization involving all available states are reported in Fig 5. Relaxations with PBE-D2 setup started either from flat monolayer (blue line) or from B1 state (cyan with triangles) definitely relax to A state. B1 appears only as an elusive state during optimization (not shown). The relaxation started from B2 (cyan with squares), conversely, stabilizes at ~+0.3 eV from A. We therefore concluded that for the PBE-D2 setup the system is definitively stabilized in the A pattern, allowing B2 state as a metastable one. With LDA setup (green lines), the behavior appears similar to PBE-D2: we first observe an energy decrease due to the readjustment of the buffer-monolayer distance (which differs in LDA and PBE-D2 setups) and then a stabilization in A type pattern, practically indistinguishable from that of PBE-D2 (inset contoured in green).

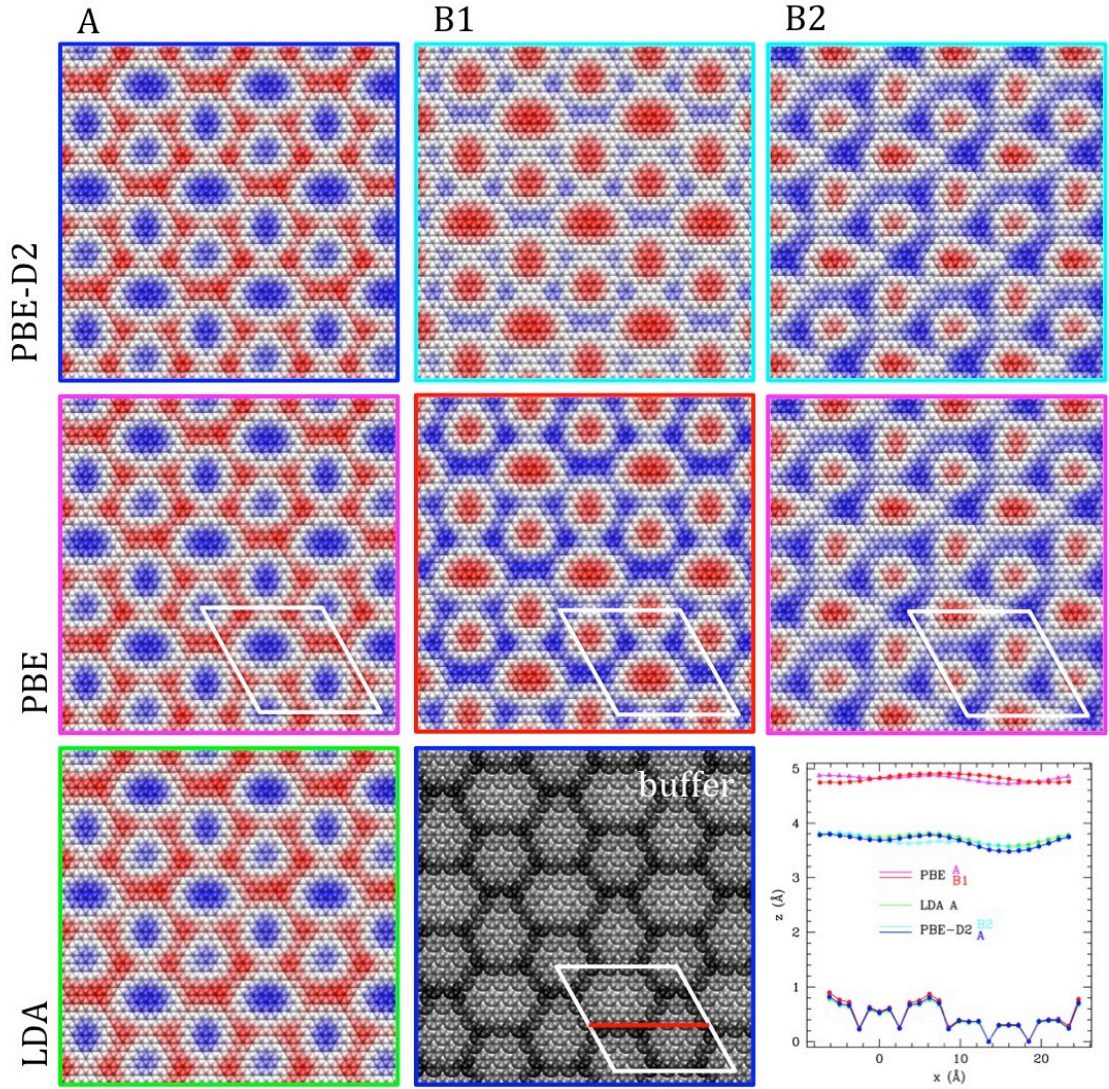

**Figure 4**. Structural details of the monolayers in the states A, B1 and B2 (first, second and third column respectively) evaluated with the PBE-D2, PBE and LDA setup (first, second and third row, respectively; for LDA only A state was evaluated). In all snapshots, the underlying buffer layer structure is aligned (reported in black and white in the last row, second column), and the supercell is reported in selected cases to show the alignment. The color coding for monolayer is red= protrusion, blue=intrusion. In the bottom-right plot the z profiles along x coordinate of the buffer and monolayer are reported for selected cases (indicated), evaluated along the red line in the buffer snapshot. Buffer layers are aligned and monolayers are located at the real z coordinate with respect to buffer.

The system behaves differently using the PBE setup. First, a noticeable buffer-monolayer distance readjustment to ~4.4Å leading to a ~7eV energy decrease (red line and data) is observed. This is due to the underestimation of the vdW interaction in this setup, also leading to overestimation of interlayer distances dominated by this interaction (see the plot in Fig 4). After this "detachment" the system definitively relaxes to B1 (also shown in Fig 3 third column). However, before the final relaxation the system explores elusive states similar to A or mixed (an "AB" like state is reported in the figure), indicating the presence of corrugation patterns with similar energy (a movie showing the corrugation variation during B1 relaxation is reports as supporting material). This is confirmed by a relaxation of A state, which results practically degenerate with B1 (just ~0.01eV more stable). In addition, the average level of curvature of all states in PBE setup is 40-50% smaller than in PBE-D2 (or LDA).

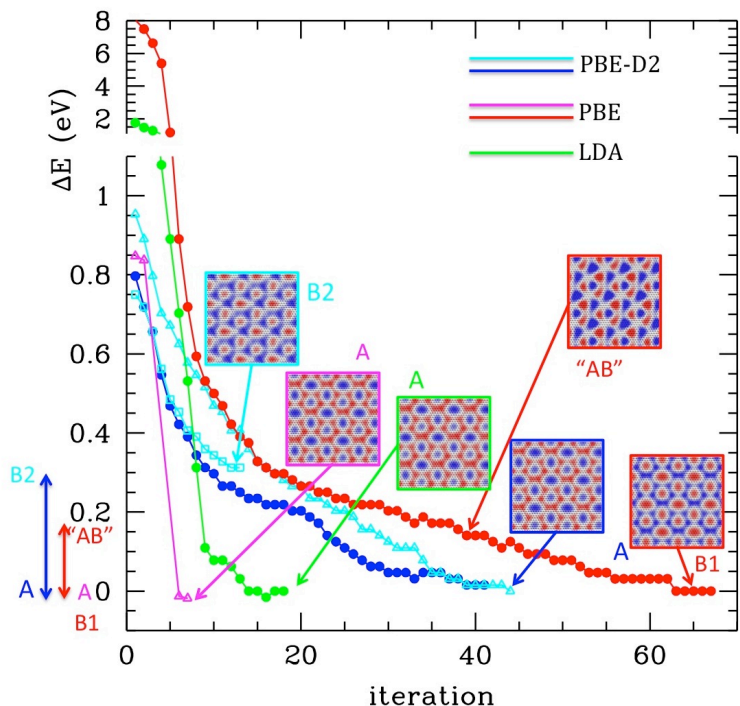

**Figure 5.** Structure relaxation runs of the "full" system with the three different calculation setup (color coding reported in the plot; double color (blue-cyan and red-magenta are used for clarity). Energies are referred to a common reference for each simulation setup. For representative states the top view of the monolayer is reported colored according to z coordinate: red=protrusion, blue=intrusion. The snapshots size ~8.6×8.6 nm$^2$, and are represented as their vdW surface. A states are characterized by blue spots (wells) contoured by red (crests), while B-like states on the contrary, by red spots (hills) contoured by blue (valleys). Energy differences between states are reported on the left. Blue data: relaxation with PBE-D2 started from flat monolayer; cyan with triangles: same started from B1-like state; cyan with squares: same started from B2-like state. Red= relaxation with PBE started from B1; magenta with triangles: same started from A. Green: relaxation with LDA setup.

The energy relations between states were further analyzed by separating elastic and vdW components. The elastic component due to the sheet corrugation was evaluated by means of a series of fixed structure calculations of the isolated sheet, taking frames along the relaxation trajectory (see The SI, section S.3.3 for details). This turns out to depend nearly quadratically on the average curvature level, defined by the root mean squared deviation average protrusion dihedral. Assuming the quadratic law holding in general, we find that the elastic component brings an energy penalty of ~0.15 and ~0.1 eV for A and B states, respectively, in PBE-D2 setup, while these values are decrease to ~0.05 eV or less for PBE states, due to their smaller curvature levels (energy values are for a single 13×13 supercell, including 338 C atoms). These energy penalties are compensated in the full model optimized structures by the vdW attraction between buffer and monolayer, which therefore has to be larger in PBE-D2 setup. In fact, our tests on graphite and graphene bilayers indicate that vdW interlayer energy is about one order of magnitude larger in PBE-D2 setup with respect to PBE, entirely due to vdW corrections.

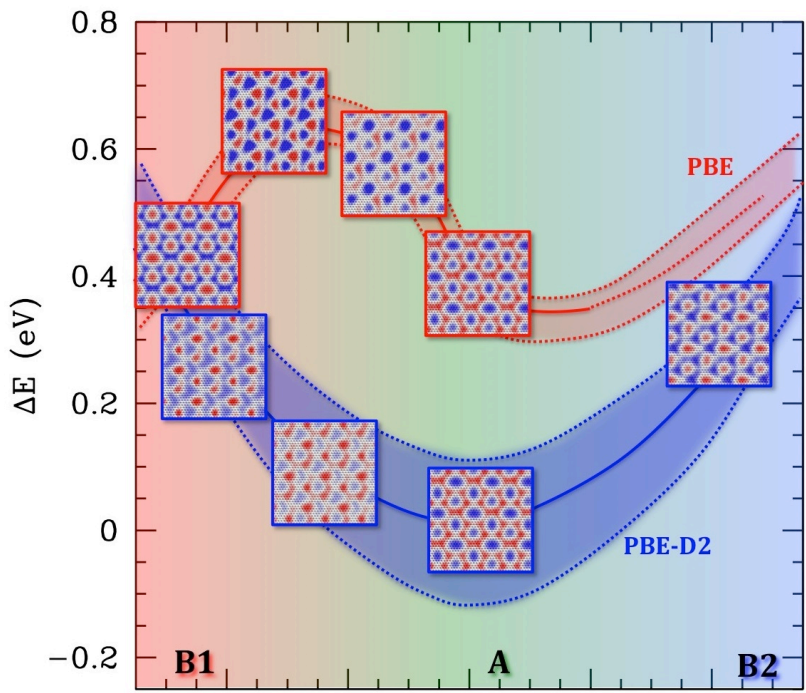

**Figure 6.** Schematic representation of the energy profiles along a reaction coordinate for the transition B1-A-B2. The structures are located at the energies where they are found along the simulations, and the blue and red lines connect them to guide the eye (for each of the two setups, the energies are arbitrarily shifted for representation convenience). The dotted lines represent the elastic component (upper lines) and vdW component (lower line) whose sum gives the solid line.

The relative energies are summarized in Fig 6. Thanks to the stronger vdW interaction, the PBE-D2 setup is more selective with respect to the monolayer corrugation states, and gives larger penalties to states B1 and B2 having the pattern not perfectly matched with the buffer; as a consequence, these are ~0.3-0.4eV less stable than A. Conversely, thanks to smaller vdW penalties, A and Bs are almost degenerate in PBE setup. We observe, in addition, that while in the PBE-D2 setup the transition between Bs and A state are spontaneous, in the case of PBE a barrier appears of ~0.1-0.2 eV.

## 4. CONCLUSIONS

This study leads to different conclusions. From the point of view of the modeling, we found that an accurate calculation setup for evaluation of corrugation of graphene on SiC must include a very accurate vdW forces representation, which are crucial to reproduce the correct structure and relative stability of corrugation patterns. *En passant*, the strong dependence of results on the level of treatment of these interactions and on the specific functional used explains the variety of different simulation results available in the literature.

We also infer that PBE-D2 setup correctly reproduces the structure and energetics at cryogenic temperatures. The buffer-monolayer distance is in agreement with experimental data, and the system appears "frozen" in the state A, predicted to be ~0.4eV more stable than B. However, the state energy differences can be overcome as the temperature rises. In particular, the thermal energy can easily fill the vdW well (estimated around 25 meV per atom in PBE-D2 in the bilayer graphene), effectively considerably reducing the vdW interaction strength. Therefore, at room temperature the energy profile might be more similar to that generated by PBE (red lines), or possibly something in between with almost degenerate A and B states and probably lower barriers in between. Accordingly, the lower level of curvature of B states with respect to A (on average 15% less) is in agreement with the smaller curvature of monolayer at room temperature recently measured [16]. This might explain and justify the use of PBE setup in the literature, in spite of its inaccuracy in reproducing some details of the structure, such as the inter-layer distance. Finally, the almost degeneracy between A and B like patterns explain why they might appear both in the STM data, especially at high temperature.

Besides the clarification of actual structure of graphene on SiC, the existence of multi-stable rippling states with similar energy and opposite/translated curvature pattern indicates that the system is prone to curvature switching controllable by environmental conditions. One of these is the temperature, which, as said, can change the relative stability of the states. Also local irregularities in the buffer layer (defects of any type) could favor one or the other curvature, and it is likely that these can be changed by e.g. doping, or external electric fields[33]. All these possibility are currently under investigation and will be matter of forthcoming papers. In turn, the possibility of controlling local curvature could allow controlling reactivity and therefore decoration and functionalization with adatoms or chemicals at the nano-scale[34], essential for a number of applications, and was proposed as the basis for gas storage and transport devices[35]. In general, these effects might be exploited in technological applications requiring the local control of graphene sheet curvature at the nanoscale.

**Supporting Information**. The following files are available free of charge: (i) A SI.pdf file with supporting data, figures and tables, all commented; (ii) Two movies files: BI_optimization.mpg, illustrating the optimization of the B1 state in PBE calculation setup and CompleteCycle_morph.mpg, a morph connecting A, B1, and B2 corrugation states.

**Corresponding Author**

valentina.tozzini@nano.cnr.it

**ACKNOWLEDGMENT**

We thank Dr Stefan Heun, Prof Paolo Giannozzi, Dr Camilla Coletti, Dr Sarah Goler and Dr Vittorio Pellegrini for useful discussions. We gratefully acknowledge financial support by the EU, 7th FP, Graphene Flagship (contract no. NECT-ICT-604391), the CINECA award "ISCRA C" IscraC_HBG, 2013 and PRACE "Tier0" award Pra07_1544 for resources on FERMI (IBM Blue Gene/Q@CINECA, Bologna Italy), and CINECA staff for technical support.